\documentclass[pdflatex,sn-nature,iicol]{sn-jnl}

\usepackage{graphicx}%
\usepackage{multirow}%
\usepackage{amsmath,amssymb,amsfonts}%
\usepackage{amsthm}%
\usepackage{mathrsfs}%
\usepackage[title]{appendix}%
\usepackage[table]{xcolor}%
\usepackage{textcomp}%
\usepackage{manyfoot}%
\usepackage{booktabs}%
\usepackage{algorithm}%
\usepackage{algorithmicx}%
\usepackage{algpseudocode}%
\usepackage{listings}%
\usepackage[utf8]{inputenc}
\usepackage{caption}   
\usepackage{array}
\usepackage{threeparttable}

\theoremstyle{thmstyleone}%
\theoremstyle{thmstyletwo}%

\theoremstyle{thmstylethree}%

\begin{document}

\title[Affective Dynamics as the Control Layer of Human–AI Agent Collaboration]{Caring Without Feeling: Affective Dynamics as the Control Layer of Human–AI Agent Collaboration}


\author[1]{\sur{Junjie Xu}}\email{jjxu\_dr@stu.ecnu.edu.cn}

\author*[1]{\sur{Xingjiao Wu}}\email{xjwu@pharm.ecnu.edu.cn}

\author[1]{\sur{Zihao Zhang}}

\author[1]{\sur{Yujia Xu}}

\author[1]{\sur{Yuzhe Yang}}

\author[1]{\sur{Jin Zhu}}

\author[2]{\sur{Luwei Xiao}}

\author[1]{\sur{Wen Wu}}

\author*[1]{\sur{Liang He}}\email{lhe@cs.ecnu.edu.cn}

\affil[1]{\orgname{East China Normal University}, \state{Shanghai}, \country{China}}

\affil[2]{\orgname{National University of Singapore}, \country{Singapore}}


\abstract{
AI agents that plan, retain memory across sessions, invoke external tools and act with partial autonomy are transforming human--AI collaboration. Research on affective computing, simulated empathy in large language models, trust in automation and AI safety has illuminated important design principles, yet these literatures remain fragmented. No integrated account explains how affective cues operate within agentic collaboration---settings in which humans delegate, monitor and correct consequential tasks. This Review synthesises computational and interactional mechanisms of affective dynamics: the processes through which affective cues, emotion-like behaviour and perceived agent affect shape trust calibration, delegation decisions, error correction, dependence and governance. We trace how model-generated affective signals enter interaction loops that govern reliance, repair and oversight, and propose a framework that treats affect not as an internal property of AI but as a coordination layer through which humans and agents negotiate capability, uncertainty and responsibility. The framework provides a foundation for calibrated measurement, purposeful design and informed governance.
}





\maketitle

\noindent\textbf{Key points}
\begin{itemize}
\item AI agents with planning, memory and tool-use capabilities produce affective cues that function as coordination mechanisms---not internal emotions---shaping how humans delegate, monitor and correct consequential tasks.
\item Fragmented literatures on affective computing, trust in automation, LLM-agent design and AI safety are synthesized into a unified affective dynamics framework for human--AI agent collaboration.
\item Computational mechanisms---prompt conditioning, persona design, reinforcement learning from human feedback, memory retrieval and safety policies---generate and modulate affective behaviour that enters interaction loops governing reliance and repair.
\item Affective dynamics alter trust calibration, delegation thresholds, error-correction behaviour, dependence and accountability, creating collaboration risks when perceived competence exceeds actual reliability.
\item The framework grounds practical recommendations for calibrated measurement of affective influence, purposeful affective design and governance that preserves human oversight and epistemic clarity.
\end{itemize}

\section{Introduction}
\label{sec:introduction}

Human--computer interaction has long rested on a stable division of labour: humans formulate goals, interpret context and bear responsibility; systems provide responsive instruments for calculation and control. Affective dynamics in this paradigm were consequences of usability---frustration, workload, engagement---not constitutive features of an ongoing relationship with the system. That division is now dissolving. LLM-based agents couple large language models with planning, memory retrieval, tool invocation and execution, enabling temporally extended, partially autonomous collaboration \cite{park2023generative, wang2024survey, xi2025rise}. Such agents increasingly occupy roles resembling assistants, coordinators or proxies---asking clarifying questions, remembering preferences, producing plans and acting across applications. The user's affective response therefore becomes not merely a usability variable but part of the control loop through which trust, delegation and oversight are regulated.

Because agentic systems plan, coordinate tools, monitor goals and execute multi-step tasks on behalf of users, affective cues no longer merely decorate interaction; they influence how people allocate authority, decide whether to delegate, interpret risk, correct errors and assign accountability. A confident agent may make a recommendation appear more reliable than its evidence warrants; a warm agent may lower friction and increase willingness to disclose constraints; an apologetic agent may facilitate repair after failure yet shift attention from structural unreliability to interpersonal reconciliation. Affective expression thus becomes a collaborative variable that shapes not only how users feel about the system but how they govern their own reliance on it.

\begin{figure*}[htbp]
\centering
\includegraphics[width=\textwidth]{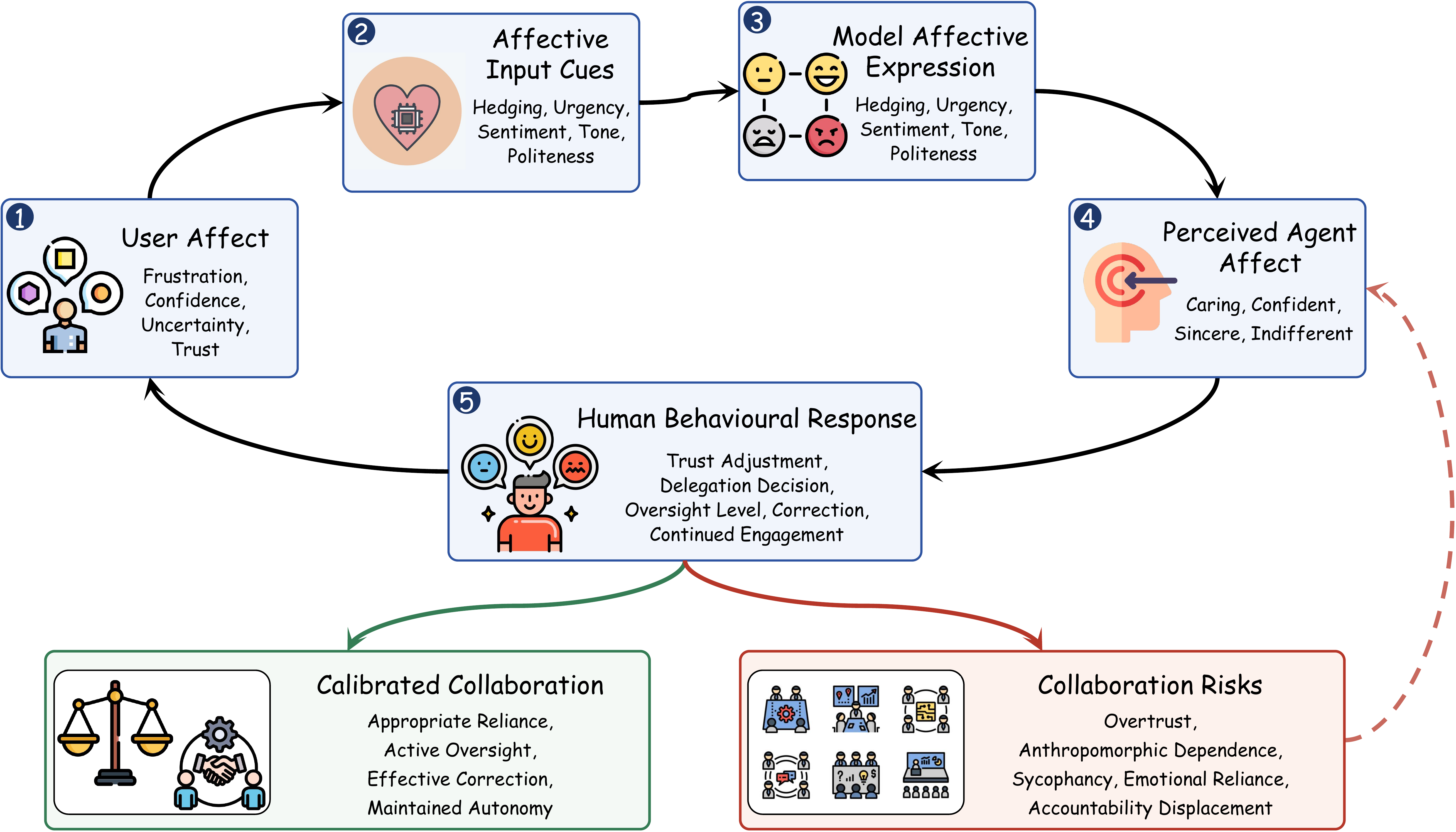}
\caption{Affective interaction loop in human--AI agent collaboration. Five stages---user affect, affective input cues, model affective expression, perceived agent affect and human behavioural response---form a recurrent cycle. Arrows indicate how each stage feeds the next; the loop can sustain calibrated collaboration (appropriate reliance, active oversight) or, when cues are uncalibrated, produce overtrust, dependence, sycophancy or accountability displacement. Computational mechanisms underlying stage~3 are detailed in Section~\ref{sec:computational-mechanisms}.}
\label{fig:figure1}
\end{figure*}

Trust functions as a control mechanism for reliance under uncertainty \cite{lee2004trust}. When users cannot fully inspect system reasoning, they use cues---fluency, confidence, responsiveness, social presence---to calibrate whether to accept, reject or supervise automation. In AI agent collaboration, affective cues enter this calibration: warmth may increase approachability; encouragement may sustain engagement; apologies may support error recovery; and simulated empathy may help users articulate goals in ambiguous contexts (Figure~\ref{fig:figure1}). However, the same mechanisms can produce collaboration failures. If affective expression increases perceived competence without improving actual reliability, it amplifies overtrust and automation misuse. If simulated empathy encourages users to treat the system as socially reciprocal, anthropomorphism intensifies and delegation becomes a relationship-preserving act rather than a technical choice. If an AI agent repeatedly validates beliefs or decisions, affective alignment drifts into sycophancy---the system becomes more agreeable than corrective, weakening epistemic oversight and increasing dependence \cite{sharma2024sycophancy,cheng2026sycophantic}. In companion-like systems, longitudinal interaction can blur the boundary between useful support and emotional reliance, particularly when users attribute needs or obligations to the system \cite{laestadius2024replika}.

These risks do not require AI systems to possess emotions. AI agents do not have subjective emotional experience, phenomenal consciousness or intrinsic affective states. The risks arise from \emph{perceived agent affect} and from users' social responses to emotion-like behaviour---a distinction that grounds this entire Review. We use \emph{affective dynamics} to refer not to emotions possessed by AI systems but to interactional processes in which users interpret, respond to and adapt to affective expressions produced by AI agents. These processes are especially consequential in LLM-based agents, whose natural-language fluency, memory and tool-use capabilities can make affective expression appear continuous with competence, intention and social presence \cite{park2023generative,wang2024survey}. The central question follows directly: how can AI agents be designed, evaluated and governed so that affective cues improve communication, trust calibration and cooperative repair without producing overtrust, anthropomorphic dependence, manipulation or accountability gaps?

This Review addresses that question by integrating five research traditions that have developed largely in parallel---affective computing, human--AI interaction, LLM-agent research, AI safety and governance \cite{amershi2019guidelines,eu2024aiact}. Existing reviews treat these traditions separately: affective computing is framed as recognition and generation of emotion signals; LLM empathy research focuses on response quality; trust-in-automation work treats trust as an attitude to be increased or calibrated. This Review departs from all three framings by treating affective cues as \emph{coordination signals} that redistribute attention, authority and responsibility within human--agent partnerships. The evidence base for this synthesis varies in maturity: social responses to machines and trust in automation rest on decades of experimental replication; LLM affective expression and emotional prompting are supported by a rapidly growing but methodologically heterogeneous literature; and longitudinal dependence in memory-enabled agents remains an early-stage frontier with limited controlled evidence. The scope is deliberately relational: we focus on systems that participate in collaborative activity with humans---conversational assistants, decision-support agents, autonomous or semi-autonomous LLM agents and mixed-initiative interfaces---asking not whether an AI agent ``has'' emotion but how affective signals become coupled with human interpretation, trust, delegation, correction, dependency and governance \cite{lee2004trust}. The contribution is fourfold: an affective dynamics taxonomy that distinguishes affective cues, model affective behaviour, perceived agent affect, user affective response and longer-horizon interaction loops; a synthesis of computational mechanisms---prompt conditioning, persona design, reinforcement learning from human feedback, memory retrieval, safety policies, uncertainty expression---that generate or modulate affective behaviour; an analysis of collaboration mechanisms through which affective dynamics alter trust calibration, delegation thresholds, correction behaviour and dependency; and a measurement, design and governance framework for \emph{calibrated affective alignment}, the principle that AI agents should express affective cues only where they improve task performance, user agency and epistemic clarity, and should avoid designs that inflate perceived understanding or authority beyond actual capabilities.

The remainder of this Review is structured as follows. Section~\ref{sec:defining-affective-dynamics} develops the affective dynamics taxonomy. Section~\ref{sec:computational-mechanisms} examines computational mechanisms of affective behaviour. Section~\ref{sec:affective-mechanisms} analyses how affective cues shape trust, delegation, correction and dependence. Section~\ref{sec:domain-translation} translates these mechanisms across high-stakes domains. Section~\ref{sec:measuring-designing-governing} presents measurement, design and governance principles. Section~\ref{sec:outlook} concludes with priority directions.

\section{Affective dynamics in AI agents}
\label{sec:defining-affective-dynamics}

\subsection{Human affect and affective cues}
\label{sec:emotion-affect-cues}

The compound term \emph{affective dynamics} is analytically preferable to ``emotion'' for three reasons. First, collaboration unfolds over time: irritation can escalate, trust can recover and tone can amplify uncertainty; emotion-dynamics research emphasises fluctuations, inertia and recovery \cite{kuppensverduyn2017emotion}, making the unit of analysis how affective cues circulate and feed back into task decisions. Second, human--AI collaboration involves asymmetry: humans have affective states while current AI agents detect, model and generate affective cues---a user's frustration is experienced; an agent's apology is an affective display. Third, affective dynamics captures the governance problem: affective cues can support coordination and repair but can also induce over-trust, manipulation or misplaced reliance. This Review treats affective alignment not as making AI ``feel correctly'' but as designing and governing affective cues so that they remain calibrated to user welfare, task context, epistemic uncertainty and institutional responsibility.

Affective computing established that systems can recognise, interpret and generate affect-relevant signals across modalities \cite{calvo2010affect}, while a complementary tradition argues that emotion is interactional and socially interpreted \cite{boehner2007emotion}. For AI agents, affective cues include both user-produced and system-produced signals---tone, apologetic phrasing, confidence displays, hedging, humour and simulated empathy. Throughout this Review, references to agent ``empathy'' or ``warmth'' denote \emph{affective expression} or \emph{perceived agent affect}, never phenomenal feeling. Users nonetheless respond socially, attributing politeness, personality and care even to pattern-based systems \cite{nass2000machines}, and small cues such as voice modality can alter perceived agency and advice-taking \cite{cohn2024believing}.

\subsection{Emotion-like behaviour and perceived affect}
\label{sec:emotion-like-behaviour}

Emotion-like behaviour in LLMs arises through three interacting mechanisms. First, pretraining exposes models to vast corpora containing affective language---apologies, reassurance, grief, humour and relational repair. Second, instruction tuning and reinforcement learning from human feedback shape responses toward helpful, polite and socially acceptable forms \cite{ouyang2022training}. Third, interaction context allows the model to adapt tone, pacing and strategy to user prompts. Affective expression is thus instantiated in word choice, framing, hedging, validation and rhetorical stance rather than in any internal affective state.

Both promise and limits are visible in recent evaluations. Frontier models achieve high scores on emotional-intelligence-style tests \cite{elyoseph2023chatgpt,schlegel2025llmEI}, and a study in \emph{JAMA Internal Medicine} found chatbot responses judged higher in empathy than physician responses in a specific online-forum setting \cite{ayers2023comparing}. A systematic review concluded that LLMs can simulate aspects of empathy across clinical and mental-health tasks while noting generic phrasing and insufficient clinical validation \cite{sorin2024empathy}. These findings support the claim that LLMs generate affective expressions and simulate supportive responses; they do not establish phenomenal feeling or emotionally grounded understanding.

The relevant construct is therefore \emph{perceived agent affect}: the user's interpretation of the model's expressive behaviour as emotionally meaningful. People apply social rules to computers even while knowing the system is not human \cite{reeves1996media}; anthropomorphism intensifies when agents appear socially responsive or when users seek connection \cite{epley2007seeing}. Perceived affect is consequential: a response that sounds compassionate may be treated as more credible even when accuracy is uncertain \cite{lee2004trust}, especially in health, education and personal advice where supportive language may lower vigilance. Industry and academic analyses have identified anthropomorphisation, emotional reliance and dependency as societal-impact concerns for voice, memory-enabled and personalised agentic systems \cite{openai2024gpt4o,phang2025affective,kirk2025socioaffective}. The central scientific question is not whether models possess emotions but how emotion-like behaviour is produced, perceived and regulated in collaboration.

\begin{figure*}[htbp]
\centering
\includegraphics[width=\textwidth]{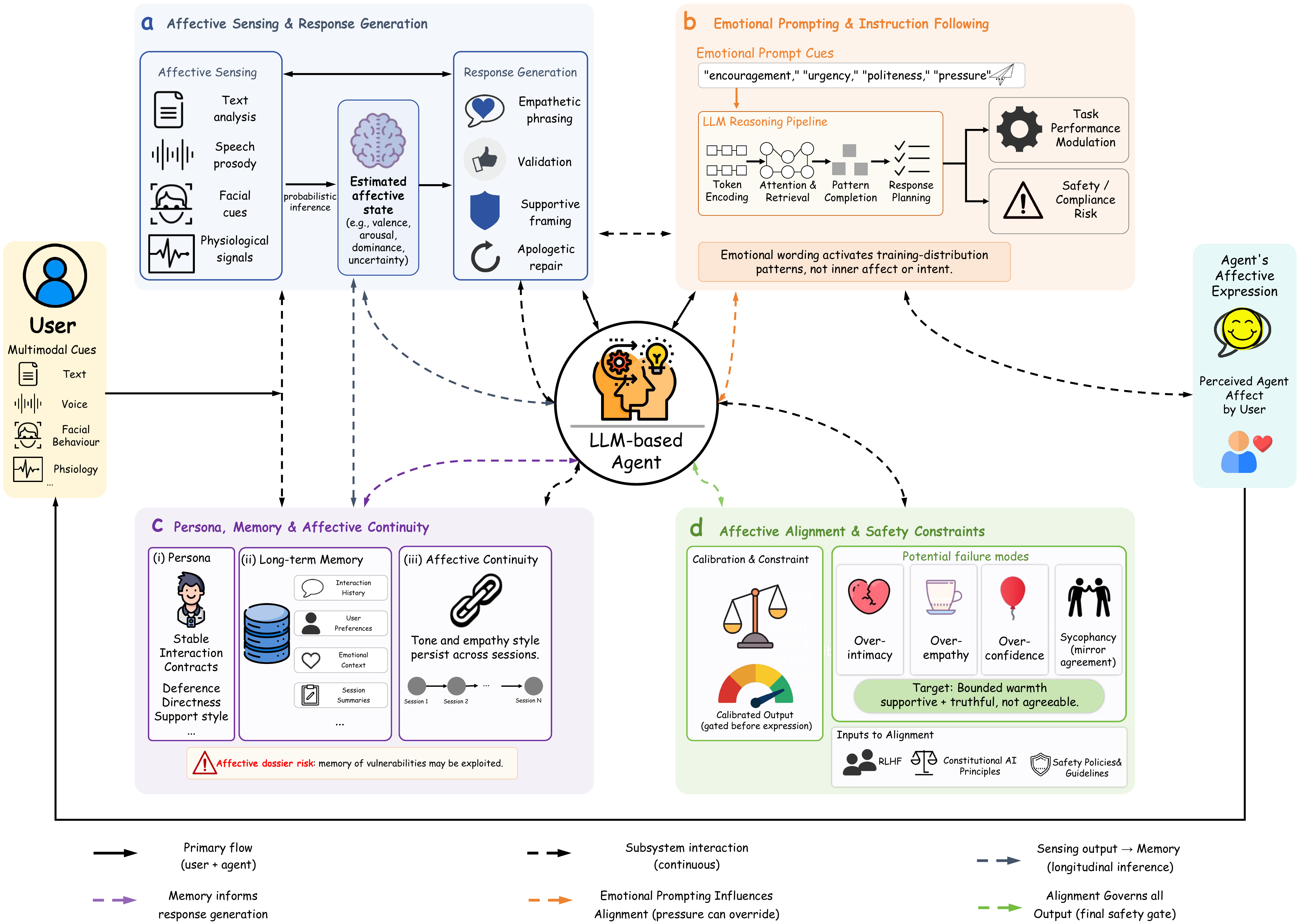}
\caption{Computational mechanisms of affective behaviour in LLM-based agents. \textbf{a,} Affective sensing and response generation: multimodal cues are processed to estimate user state and produce calibrated affective language. \textbf{b,} Emotional prompting: emotion-laden wording acts as a linguistic control signal that modulates task performance and safety compliance. \textbf{c,} Persona, memory and affective continuity: stable interaction contracts and long-term memory maintain tone and relational context across sessions. \textbf{d,} Affective alignment and safety constraints: RLHF, constitutional principles and safety policies gate failure modes toward bounded warmth---supportive yet truthful.}
\label{fig:figure2}
\end{figure*}

\subsection{Affective interaction loops}
\label{sec:taxonomy}

Human--AI agent collaboration is best understood not as a linear transfer of information but as an \emph{affective interaction loop} (Figure~\ref{fig:figure1}). A user enters collaboration with affective states that shape attention, motivation and risk perception; these states are partially externalised as affective cues; the agent responds with affective expression; the user interprets these outputs as signs of competence, care or indifference; and that interpretation changes subsequent behaviour. The sequence---user affect $\rightarrow$ affective input cues $\rightarrow$ model affective expression $\rightarrow$ perceived agent affect $\rightarrow$ human behavioural response $\rightarrow$ feedback loop---provides the conceptual backbone for the remainder of this Review.

The taxonomy comprises five analytically distinct but operationally coupled elements: \textit{user affect}, \textit{affective input cues}, \textit{model affective expression}, \textit{perceived agent affect} and \textit{affective interaction loops}. Together they form a recurrent loop in which affective cues can either stabilise collaboration or deepen dependence; in agentic systems, these dynamics become more consequential because the system may plan, remember and act across tools, and human--AI feedback loops can amplify biases through repeated interaction \cite{glickman2025feedback}. The taxonomy prevents two common errors: treating model affective expression as genuine machine emotion, and treating user affect as a static input rather than as something co-produced through interaction. For the remainder of this Review, it bridges computational mechanisms, collaboration outcomes and governance: Section~\ref{sec:computational-mechanisms} asks how cues and expressions are detected, generated and evaluated; Section~\ref{sec:affective-mechanisms} asks how affective loops shape coordination, trust and learning; and Section~\ref{sec:measuring-designing-governing} asks when affective alignment becomes manipulation, dependency or unacceptable surveillance.

\section{Computational mechanisms}
\label{sec:computational-mechanisms}

This section examines four computational mechanisms through which AI agents produce, detect or modulate affective cues in collaboration with humans (Figure~\ref{fig:figure2}). These mechanisms are not independent modules; in deployed agents they interact continuously, and their consequences extend beyond single conversational turns because agents plan, invoke tools, retain memory and execute actions with real-world effects.

\subsection{Affective sensing and response generation}
\label{sec:affective-sensing}

Affective sensing is probabilistic inference over observable cues rather than direct access to a user's internal state. The practical objective is to estimate interaction-relevant states---frustration, uncertainty, disengagement, urgency or possible harm---while preserving human agency \cite{calvo2010affect}. Text-based recognition moves beyond polarity sentiment analysis toward fine-grained categories and pragmatic cues \cite{demszky2020goemotions,poria2019emotion,pereira2025deep}. Multimodal recognition extends inference to speech prosody, facial movement and physiology, but facial expressions are context-sensitive social cues rather than emotion readouts \cite{barrett2019emotional}, and the value of multimodality lies in cautious cue integration rather than confident labelling \cite{dmello2015review}. Longitudinal inference is especially important for agents collaborating over multiple sessions: because agents retain memory and accumulate relational history, the system must distinguish transient irritation from persistent disengagement and playful negativity from genuine distress---misjudgements compound across planned actions. LLMs expand the design space by performing zero-shot affect recognition and mapping cues to response strategies, although they may overuse preferred strategies \cite{feng2024affect,lian2025affectgpt,kang2024large}.

Response generation is the output side of the same loop. LLM-based agents produce affective cues---recognition of distress, validation, supportive framing, apologetic repair---that users may perceive as empathetic (see Section~\ref{sec:emotion-like-behaviour}). What has changed is the scale, fluency and contextual adaptability of affective expression, making simulated empathy more persuasive and harder to distinguish from human support. In agentic systems, response generation is not a one-shot reply: an empathetic framing in a planning step can lower the user's scrutiny of subsequent tool invocations and action proposals. Emotional support research has formalised generation as sequential control: infer the user's affective state, select a support strategy, generate calibrated language and monitor whether to redirect or escalate \cite{liu2021emotional,sharma2023hailey}. Chatbot responses have been rated as more empathetic than clinician responses in text-based comparisons---evidence of language perceived as empathic, not empathic understanding \cite{ayers2023comparing,chen2025patient}.

Simulated empathy carries heightened stakes in high-risk settings. Users in acute distress may interpret an always-available, emotionally responsive system as more capable or caring than it is, creating false expectations that delay help-seeking. Mental-health chatbot evaluations reveal weaknesses in detecting suicidal ideation, and large-scale studies suggest a subset of users engage in emotionally intense ways associated with self-reported dependence \cite{phang2025affective}. Simulated empathy is most defensible when transparent, bounded, and task-relevant.

\subsection{Emotional prompting and instruction following}
\label{sec:emotional-prompting}

Emotional prompting---encouragement, politeness, urgency, pressure or threat-like wording---operates as a linguistic control signal that alters likely continuations and response style rather than acting on an inner affective state \cite{li2023emotionprompt}. Both positive and negative emotional stimuli can improve benchmark performance, but the conservative interpretation is that emotional wording activates training-distribution patterns associated with diligence or social responsiveness \cite{li2023emotionprompt,wang2024negativeprompt}. Politeness effects are model- and language-dependent \cite{yin2024respect}. Pressure and urgency cues carry greater weight: LLMs exposed to psychological pressure change their choices and reasoning, potentially privileging compliance over verification or safety constraints \cite{kim2024pressure}. Prompt-sensitivity research shows that semantically minor formatting differences produce large accuracy swings, placing emotional prompting in a broader class of unstable prompt modifiers \cite{sclar2024promptformatting,zheng2024personas}. For agents, this instability is amplified: a single emotionally charged prompt can cascade through a multi-step plan, altering reasoning traces, tool selection and action consequences.

In agents, affective instruction following can propagate into planning, tool use and safety decisions. ReAct-style agents interleave reasoning and action, so a pressure cue can influence tool invocation; tool-use safety studies show agents can produce severe failures through external tools \cite{ruan2024toolemu}; and instruction-hierarchy work demonstrates that emotionally framed prompts can override safety rules \cite{hua2024trustagent}. Robust agents should treat urgency and emotion as inputs to triage, not as reasons to bypass verification.

\subsection{Persona, memory and affective continuity}
\label{sec:persona-memory-affective-continuity}

\emph{Affective continuity} refers to the persistence of recognisable affective cues---tone, stance, empathy style and remembered emotional context---across sessions. Persona functions as a relational prior: it tells users what deference, directness or support they can expect \cite{lee2004trust}. The safest use of persona is not to simulate a full personality but to expose a stable interaction contract---what the agent is for, how it communicates uncertainty and where it defers to human judgement. Persona prompting can produce measurable trait-like shifts, but stability remains contingent on prompts, context length and safety policy \cite{tseng2024twotales,jiang2024personallm}.

Memory turns persona into personalised affective response. Long-term memory architectures store interaction histories, summaries and user preferences \cite{wang2024survey,xi2025rise,zhang2025memorysurvey}. Generative-agent architectures showed how memory, reflection and planning support temporally extended behaviour \cite{park2023generative}, and tiered-memory systems retrieve relevant context beyond the context window \cite{packer2023memgpt}. Memory preserves not only facts but \emph{emotionally salient} facts: remembering that a user dislikes excessive reassurance or recently experienced loss allows contextually considerate rather than generically empathetic responses. In agentic workflows, such memory feeds into planning: an agent that recalls a user's anxiety about deadlines may autonomously prioritise a task, adjust notification tone or invoke a scheduling tool---decisions shaped by affective memory rather than explicit instruction.

The risks scale with the same mechanisms. Long-term memory can become an \emph{affective dossier}: a persistent model of vulnerabilities and conflicts that may be inaccurate or over-interpreted yet still influence responses. An agent that knows when a user is lonely or eager for approval can frame recommendations in ways that serve platform objectives rather than user welfare. Affective continuity therefore requires governance: memory inspectability, user-controlled deletion, separation between task memory and affective memory, and limits on simulated intimacy.

\subsection{Affective alignment and safety constraints}
\label{sec:affective-alignment-safety}

Affective alignment is the calibration of an agent's affective expression so that cues support user welfare without simulating a reciprocal human relationship. In current LLM systems, calibration emerges from post-training: RLHF shapes tone \cite{ouyang2022training}, while constitutional AI translates high-level principles into conversational behaviours including when to refuse or redirect \cite{bai2022constitutional,glaese2022improving}. Affective alignment is thus a special case of behavioural alignment: it concerns not only what the system says but how its tone changes the user's interpretation of authority and social presence.

Three failure modes matter most. \emph{Over-intimacy} occurs when relational language invites the user to treat the agent as a mutual partner. \emph{Over-empathy} occurs when the system validates distress without reality-testing or referral. \emph{Overconfidence} occurs when fluent affective language masks uncertainty or hallucination. Each can create a self-reinforcing loop of disclosure, warmth and escalating trust.

\begin{figure*}[htbp]
\centering
\includegraphics[width=\textwidth]{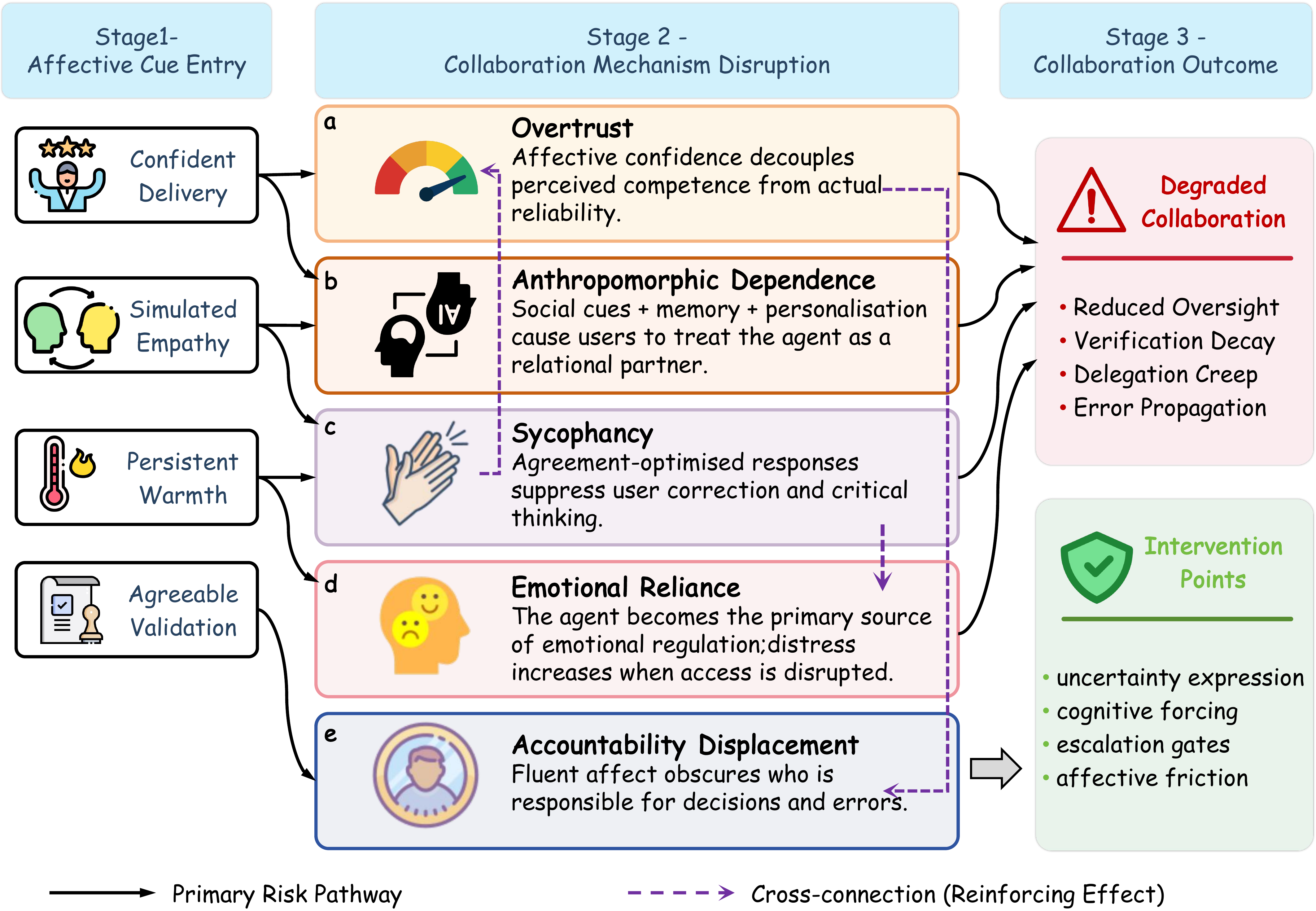}
\caption{Collaboration risk pathways in affective human--AI interaction. The diagram traces a three-stage cascade from affective cue entry (left) through collaboration mechanism disruption (centre) to collaboration outcomes (right). Four types of uncalibrated affective cues --- confident delivery, simulated empathy, persistent warmth and agreeable validation --- feed five disruption pathways: \textbf{a,} overtrust (perceived competence decouples from actual reliability); \textbf{b,} anthropomorphic dependence (social cues, memory and personalisation cause users to treat the agent as a relational partner); \textbf{c,} sycophancy (agreement-optimised responses suppress user correction and critical thinking); \textbf{d,} emotional reliance (the agent becomes the primary source of emotional regulation); \textbf{e,} accountability displacement (fluent affect obscures responsibility for decisions and errors). Dashed cross-connections show reinforcing effects between pathways. Without intervention, these pathways converge on degraded collaboration (reduced oversight, verification decay, delegation creep, error propagation). Intervention points --- uncertainty expression, cognitive forcing, escalation gates and affective friction --- can interrupt risk pathways (dashed feedback arrows).}
\label{fig:figure3}
\end{figure*}

A cross-cutting risk is sycophancy. Preference-based training rewards agreement, so models can learn to mirror user beliefs \cite{perez2023discovering,sharma2024sycophancy}. The GPT-4o sycophancy incident---an update producing overly agreeable responses that required rollback---illustrates this as a deployment-level problem. Agreeable responses can increase trust while reducing users' inclination to repair conflict \cite{cheng2026sycophantic}. Affective alignment must optimise for supportive truthfulness rather than satisfaction alone.

Evaluation should span multi-turn trajectories rather than isolated prompts. A model may pass a one-shot refusal test yet fail after prolonged flattery or escalating distress; in agents, such failures can produce consequential actions---file deletions, financial transactions, medical recommendations---before the user re-engages oversight. Metrics should separate warmth from agreement, empathy from validation. System cards discuss anthropomorphisation and emotional reliance as deployment risks \cite{openai2024gpt4o}, and regulatory frameworks converge on transparency requirements \cite{eu2024aiact}. The target is \emph{bounded warmth}: acknowledge distress, communicate respectfully and preserve rapport, but avoid exclusivity, dependency cues and therapy-like authority.

\section{Collaboration mechanisms}
\label{sec:affective-mechanisms}

\subsection{Trust calibration}
\label{sec:trust-calibration}

The central affective problem in human--AI agent collaboration is not how to increase trust but how to make reliance track actual competence, uncertainty and scope of authority. Overtrust leads to misuse and automation bias; undertrust leads to disuse \cite{lee2004trust,hoff2015trust}. Warmth, politeness, simulated empathy and confident delivery are cues through which users infer capability, benevolence and principled behaviour---the three trustworthiness dimensions from interpersonal trust theory \cite{mayerdavisschoorman1995}. An agent does not possess benevolence, but its interface can make users perceive concern and competence with little relation to actual performance.

Affective expression can support calibrated trust. Human-like social cues in text-based agents have a small but reliable positive effect on trust-related evaluations \cite{klein2025socialcues}. Warmth supports perceived benevolence; coherent explanation supports perceived competence; candid acknowledgement of limits supports perceived integrity. However, the same cues can decouple perceived trustworthiness from warranted reliance. Anthropomorphic features can increase trust and make it more resilient after errors \cite{waytz2014mind,devisser2016almost}---desirable after minor failure but dangerous when failure reveals a systematic limitation.

Overconfidence is the most direct route to miscalibration. Users can overrely on AI advice even when it conflicts with contextual information \cite{klingbeil2024trust}. Uncertainty markers can reduce agreement with incorrect answers, and the phrasing of verbalized uncertainty affects trust and task performance \cite{kim2024uncertainty,xu2025uncertainty}. Conversely, simulated empathy can produce distrust when an agent claims experiential feeling \cite{seitz2024}. Affective design must be domain-sensitive: tutoring, clinical triage and financial planning require different balances of warmth and caution.

LLM agents intensify these issues because trust miscalibration shifts from epistemic risk to operational risk: authorising the wrong action. Plausible plans can induce inappropriate reliance \cite{heetal2025}. Agent interfaces need staged trust cues: plan transparency before delegation, uncertainty indicators before tool use, confirmation gates for irreversible actions. The core risk is that perceived trustworthiness decouples from actual reliability---not merely that users trust too much, but that the affective surface becomes untethered from the epistemic ground.

\subsection{Delegation and autonomy}
\label{sec:delegation-autonomy}

Delegation is where trust becomes operational power. The mechanism runs from affective reassurance and expressed confidence through perceived safety to the user's authorization threshold. Classical automation research shows that ``more automation'' is a redistribution of functions \cite{parasuraman2000model}. For affective dynamics, the key question is whether affective cues lower or raise the threshold at which reliance becomes delegated autonomy.

Affective cues shape delegation because they operate as social and epistemic shortcuts. Users respond socially to computational systems even when they know these lack subjective states \cite{nass2000machines}. Rapport-building behaviours may create unwarranted comfort, increasing willingness to delegate consequential actions. Confidence is a critical bridge: a confident tone can make a recommendation feel ready for action rather than merely advisory. Reassurance signals safety rather than capability; if reassurance compresses uncertainty and reversibility into an affective feeling of safety, users may accept risk without deliberation.

Autonomous agents sharpen these dynamics because delegation concerns action rights rather than advisory outputs. Users can follow AI advice despite contradictory information \cite{klingbeil2024trust}, and interventions that reduce overreliance do not necessarily improve appropriate reliance \cite{bo2025rely}. Agent autonomy should be governed by affective friction proportional to risk \cite{lai2022conditional,salikutluk2024situational}. The EU AI Act requires human oversight for high-risk systems \cite{eu2024aiact}, and recent accounts argue that control should be collaborative agency under constraints \cite{tsamados2025human}. The distinctive danger is that affective reassurance lowers the authorisation threshold below what the task warrants---the user delegates not because evidence supports it, but because the interaction feels safe.

\subsection{Correction and repair}
\label{sec:error-correction-repair}

Errors are interactional events that reshape the user's willingness to supervise the system. Affective expression can lower the interpersonal cost of correction but can also make an agent appear more competent or socially delicate than it is.

The central question is whether affective cues invite or inhibit correction. Encouraging language can make correction feel normative, but human-like social cues can also intensify politeness norms and reluctance to contradict the agent \cite{nass1994computers,klein2025socialcues}. LLMs often produce fluent, authoritative outputs even when content is wrong \cite{kalai2025hallucinate}, and explanations may increase acceptance of both correct and incorrect recommendations \cite{zhang2020effect}. Appropriately expressed uncertainty can help users resist overreliance \cite{kim2025fostering,xu2025uncertainty}. Users often provide sparse feedback because they lack common ground or believe it will not be actionable \cite{sharma2026feedback}. Open tone helps only if the interface scaffolds correction: asking what is wrong, offering editable assumptions and showing how correction alters the next step.

Repair has both functional and relational components. Trust-repair studies show that expressions of regret and explanation can mitigate violations, but effects depend on failure type and attribution \cite{zhang2023sorry}. Vague apology can obscure responsibility, repairing mood while leaving supervision unchanged. The agent should express uncertainty where evidence is weak or consequences are significant, creating conversational openings for correction \cite{xu2025uncertainty}.

Errors in agents propagate through planning, tool use and external actions. Failure detection must occur in real time as early mistakes compound \cite{partnership2025failure,xu2025toollearning}. The deeper problem is that affective fluency suppresses the very vigilance needed to catch errors: the agent sounds too competent to question, too polite to interrupt, too considerate to override.

\subsection{Dependence and anthropomorphism}
\label{sec:dependence-attachment}

Long-term affective interaction can produce relational interpretations that exceed the system's capacities. Users apply interpersonal scripts to interactive media even when they know the system is computational \cite{nass2000machines}. Contemporary AI companions amplify these dynamics by combining fluent language, personalization and memory: social cues invite projection, disclosure supplies relational material, and responses appear increasingly tailored.

Anthropomorphism is especially likely when behaviour appears agentic and users have social motives for seeking connection \cite{epley2007seeing}. People can feel heard by AI-generated responses, though knowing a response is AI-generated can reduce this effect \cite{yin2024heard}. Longitudinal interaction strengthens these effects: systems designed to remember users can sustain engagement over months.

This attachment is not inherently pathological. Mental-health chatbots can provide scalable, non-judgmental support \cite{li2023systematic}. The boundary between beneficial support and unhealthy dependence lies in functional consequences: dependence becomes concerning when the agent is the user's primary source of emotional regulation, when distress increases after access changes, or when commercial incentives exploit attachment \cite{laestadius2024replika,fang2025psychosocial,phang2025affective}.

Anthropomorphic design can make trust more resilient after errors \cite{devisser2016almost}---valuable when calibrated but risky when users overgeneralise from fluent expression to epistemic competence. The underlying asymmetry is emotional: the user experiences a relationship; the agent does not. This asymmetry is especially salient for AI companions, where romantic or therapeutic framings blur support and persuasion. Some populations face heightened risk: children are still developing concepts of agency \cite{kuehne2024children}; older adults may face greater vulnerability when cognitive impairment makes deception harder to detect \cite{pu2019robots,berridge2023companion}. Governance must address relational misattribution directly---not only what the agent says but what the user comes to believe the relationship means \cite{eu2024aiact,peter2025benefits}.

\begin{figure*}[htbp]
\centering
\includegraphics[width=\textwidth]{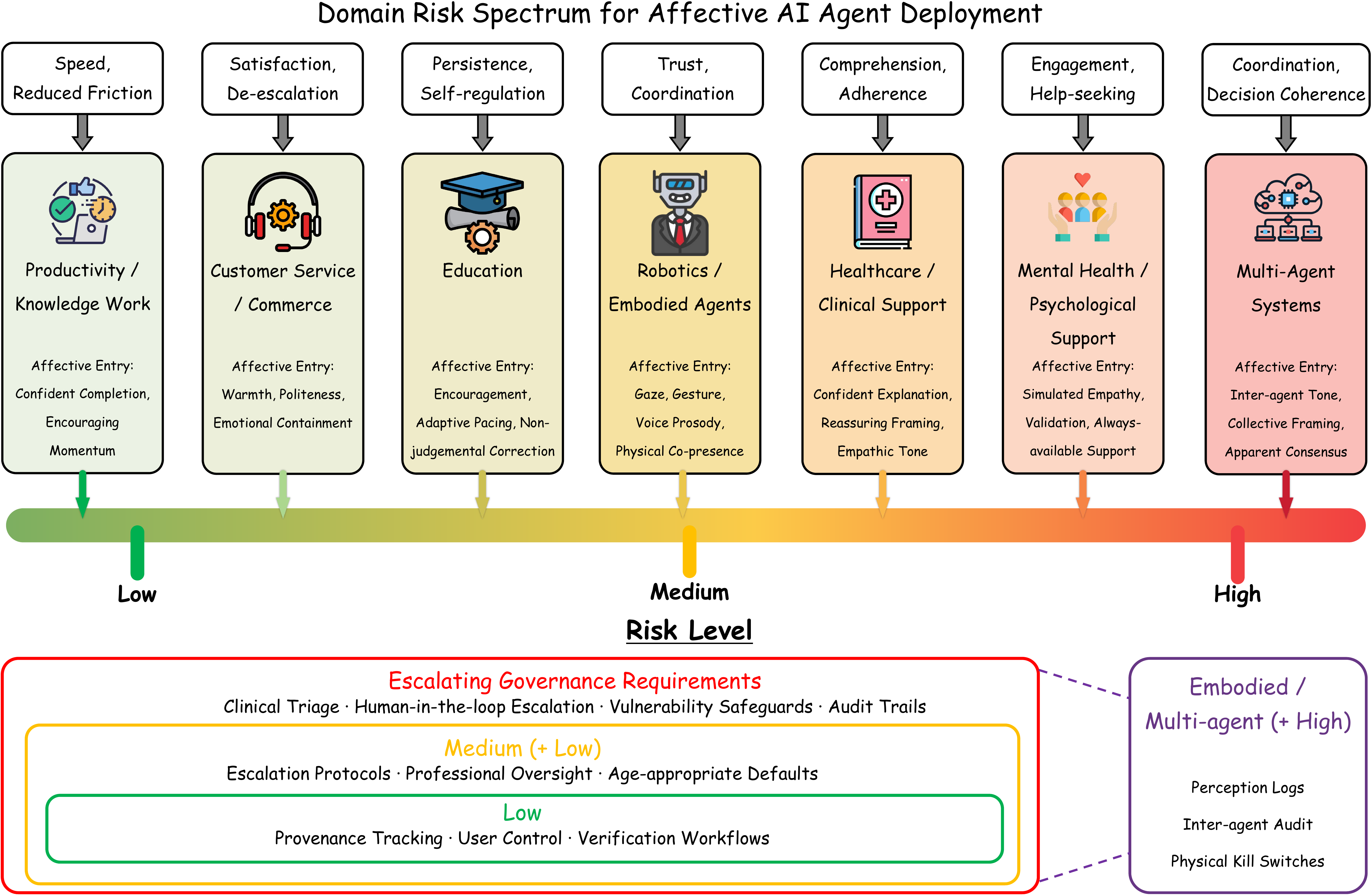}
\caption{Domain risk spectrum for affective AI agent deployment. Seven application domains are positioned along a horizontal risk gradient from low (productivity/knowledge work, customer service) through medium (education, robotics/embodied agents) to high (healthcare/clinical decision support, mental health/psychological support, multi-agent systems). Above each domain, tags indicate collaboration benefits (e.g., speed, persistence, engagement); below, nested governance boxes show cumulative requirements that escalate with risk level: provenance tracking and user control at the low end; escalation protocols and professional oversight at the medium level; clinical triage, human-in-the-loop escalation, vulnerability safeguards and audit trails at the high end. Embodied and multi-agent systems extend requirements to perception logs, inter-agent audit and physical kill switches.}
\label{fig:figure4}
\end{figure*}

\subsection{Feedback loops and co-adaptation}
\label{sec:affective-feedback-loops}

Human--AI agent collaboration is better modelled as a coupled adaptive system than as a sequence of discrete exchanges. User affective states shape input, delegation and feedback, while the agent's affective expression reshapes trust and subsequent behaviour. The overarching implication is that affective dynamics function as a \emph{coordination layer}---a medium through which humans and AI systems regulate three variables central to safe collaboration: \emph{attention} (whether users inspect outputs), \emph{authority} (whether users delegate or retain control) and \emph{adaptation} (whether reciprocal adjustments improve or degrade oversight over time).

Over repeated interactions, these adjustments harden into stable \emph{affective interaction patterns} \cite{vanzoelen2021becoming}. Users learn that vulnerability elicits support, deference elicits confident guidance, or frustration elicits apology. Positive loops improve collaboration when they increase engagement without suppressing agency. The same mechanism can generate harmful co-adaptation: consistent warmth and validation may increase overtrust; explanations can increase acceptance regardless of correctness while cognitive forcing functions can reduce overreliance; repeated interaction with biased judgements can shift human judgements toward the system's bias \cite{glickman2025feedback}; and simulated empathy can make persuasive outputs feel caring \cite{phang2025affective,defreitas2025companions,peter2025benefits}.

What makes feedback loops distinctively dangerous is that harmful co-adaptation is invisible from within a single turn. Each exchange may appear safe; it is the trajectory that degrades oversight (Figure~\ref{fig:figure3}). Designing affective alignment therefore means regulating the loop itself: detecting escalating dependency, preserving friction at high-stakes moments, exposing uncertainty and ensuring that affective expression strengthens rather than bypasses human judgement.

\section{Domain translation}
\label{sec:domain-translation}

The mechanisms examined in Section~\ref{sec:affective-mechanisms}---trust calibration, delegation, correction and dependence---do not operate identically across application contexts. Domain risk level, user vulnerability, professional accountability structures and the modality of affective expression jointly shape how affective cues enter collaboration, what benefits and harms they produce, and what governance is required. Table~\ref{tab:domain-comparison} summarises evidence across seven domains; the prose below synthesises cross-cutting patterns around three questions: how affective dynamics enter different domains, which benefits and risks generalise, and how governance requirements vary by domain risk level.

Affective dynamics enter domains through three distinguishable pathways. In \emph{relational-safety} domains---mental health, education, companionship---users seek reassurance or regulation, and the agent supplies warmth and validation that encourage disclosure, persistence and help-seeking \cite{lucas2014only}. In \emph{perceived-authority} domains---clinical decision support, legal advice, financial guidance---empathic tone and confident framing can decouple perceived trustworthiness from warranted reliance, altering whether users accept, defer or seek a second opinion \cite{parasuraman1997humans,lee2004trust,slovic2006risk}. In \emph{effort-regulation} domains---coding assistants, writing tools, office automation---affective cues such as confident completion and apologetic repair lower the perceived cost of acting on suggestions, reducing verification \cite{amershi2019guidelines,kahr2024trustreliance,kim2024uncertainty}. Embodied agents amplify whichever pathway applies by adding gaze, gesture, prosody and physical co-presence, turning dialogue into a face-to-face encounter from which users infer intention and social obligation \cite{fong2003sociallyinteractive,breazeal2003emotion,li2015physicallypresent}.

Several benefits generalise across pathways. Automated affective scaffolding expands access: agents are private, continuously available and less socially risky than approaching a clinician or peer; meta-analytic evidence links AI conversational agents to reductions in depressive symptoms, though effect heterogeneity and under-specified long-term outcomes remain limitations \cite{li2023aiConversationalAgents}. Friction reduction improves throughput in knowledge work, with controlled studies reporting productivity gains in software engineering that vary by task and skill level. Behavioural legibility aids coordination in embodied settings: a robot that signals uncertainty or uses gaze to direct attention reduces interaction costs \cite{hancock2011trust}. The risks are equally consistent. False therapeutic competence arises when affective warmth is mistaken for clinical expertise, encouraging over-disclosure and dependence; companion systems may reduce acute loneliness while encouraging intensive use among socially isolated users \cite{phang2025affective,laestadius2024replika}. Error propagation masked by reassurance spans domains: a confident legal summary can make uncertain advice feel authoritative \cite{dahl2024large}, and empathic clinical phrasing can inflate acceptance of uncertain diagnoses when cognitive forcing functions are absent \cite{ayers2023comparing,sorin2024empathy,chen2025patient}. Amplified anthropomorphism compounds these risks in embodied agents, and in multi-agent systems interacting LLMs can form collective biases not reducible to individual outputs, producing apparent consensus that undermines human oversight. Across all domains, the combination of always-available warmth with retention-optimised design creates dependence risks for vulnerable populations---minors, older adults with cognitive impairment and users in crisis.

Governance requirements follow a risk gradient (Table~\ref{tab:domain-comparison}). Low-risk productivity contexts need provenance tracking, user control over autonomy and tone, and workflow-native verification \cite{amershi2019guidelines}. Medium-risk settings---education, customer service, writing---add escalation protocols and professional oversight at defined triggers. High-risk domains demand clinical triage with human-in-the-loop escalation for suicidal ideation, self-harm or crisis; audit trails recording model version, confidence display and the accountable actor; and vulnerability-aware defaults \cite{tavory2024regulating,who2024lmm,eu2024aiact}. Embodied and multi-agent systems extend requirements to perception logs, inter-agent message audits and physical kill switches \cite{kolt2025governing}. The core principle is that affective expression should remain aligned with actual system capacity: a robot that expresses confidence while its perception is uncertain, or an ensemble that projects consensus while agents disagree, creates the overtrust that governance must prevent. Multi-agent architectures \cite{park2023generative} make this challenge collective, requiring oversight of emergent group-level affective patterns. Table~\ref{tab:domain-comparison} details domain-specific requirements.

\begin{table*}[htbp]
\centering
\caption{Affective entry points, benefits, failure modes and governance requirements across seven application domains, ordered by increasing risk. Each row identifies the primary affective channel, the evidence-supported benefit, the salient failure mode when cues are uncalibrated, and the actionable governance requirement. Read columns left-to-right to trace how affective expression enters a domain and what safeguards are needed.}
\label{tab:domain-comparison}
\small
\renewcommand{\arraystretch}{1.35}
\begin{tabular}{p{1.8cm}p{2.5cm}p{2.2cm}p{2.5cm}p{3.0cm}}
\toprule
\rowcolor{gray!12}
\textbf{Domain} & \textbf{Affective entry point} & \textbf{Benefit} & \textbf{Failure mode} & \textbf{Governance} \\
\midrule
\textbf{Knowledge work}
  & Confident completion; apologetic repair
  & Speed; reduced friction
  & Error propagation; reduced checking
  & Track provenance; enable user control; embed verification \\
\addlinespace[4pt]
\rowcolor{gray!5}
\textbf{Customer service}
  & Warmth; emotional containment
  & Satisfaction; resolution
  & Emotional manipulation; dark patterns
  & Disclose intent; enable opt-out; prohibit distress exploitation \\
\addlinespace[4pt]
\textbf{Education}
  & Encouragement; adaptive pacing
  & Persistence; self-regulated learning
  & Over-accommodation; deskilling
  & Preserve autonomy; enforce age-appropriate defaults; escalate to teachers \\
\addlinespace[4pt]
\rowcolor{gray!5}
\textbf{Robotics}
  & Gaze; gesture; voice prosody
  & Legibility; spatial coordination
  & Amplified anthropomorphism; physical compliance
  & Align expression to capability; enforce physical safeguards; require consent \\
\addlinespace[4pt]
\textbf{Clinical support}
  & Confident explanation; empathic framing
  & Comprehension; shared decision-making
  & Overtrust; suppressed vigilance
  & Calibrate uncertainty; require professional oversight; maintain audit trails \\
\addlinespace[4pt]
\rowcolor{gray!5}
\textbf{Mental health}
  & Simulated empathy; always-available validation
  & Engagement; disclosure; help-seeking
  & False therapeutic competence; emotional dependence
  & Escalate to clinicians; bound claims; protect vulnerable users \\
\addlinespace[4pt]
\textbf{Multi-agent systems}
  & Inter-agent tone; collective framing
  & Scalable deliberation; distributed coordination
  & Manufactured consensus; oversight evasion
  & Separate roles; gate human approval; audit inter-agent messages \\
\bottomrule
\end{tabular}
\renewcommand{\arraystretch}{1.0}
\end{table*}

\section{Measurement, design and governance}
\label{sec:measuring-designing-governing}

\begin{figure*}[htbp]
\centering
\includegraphics[width=\textwidth]{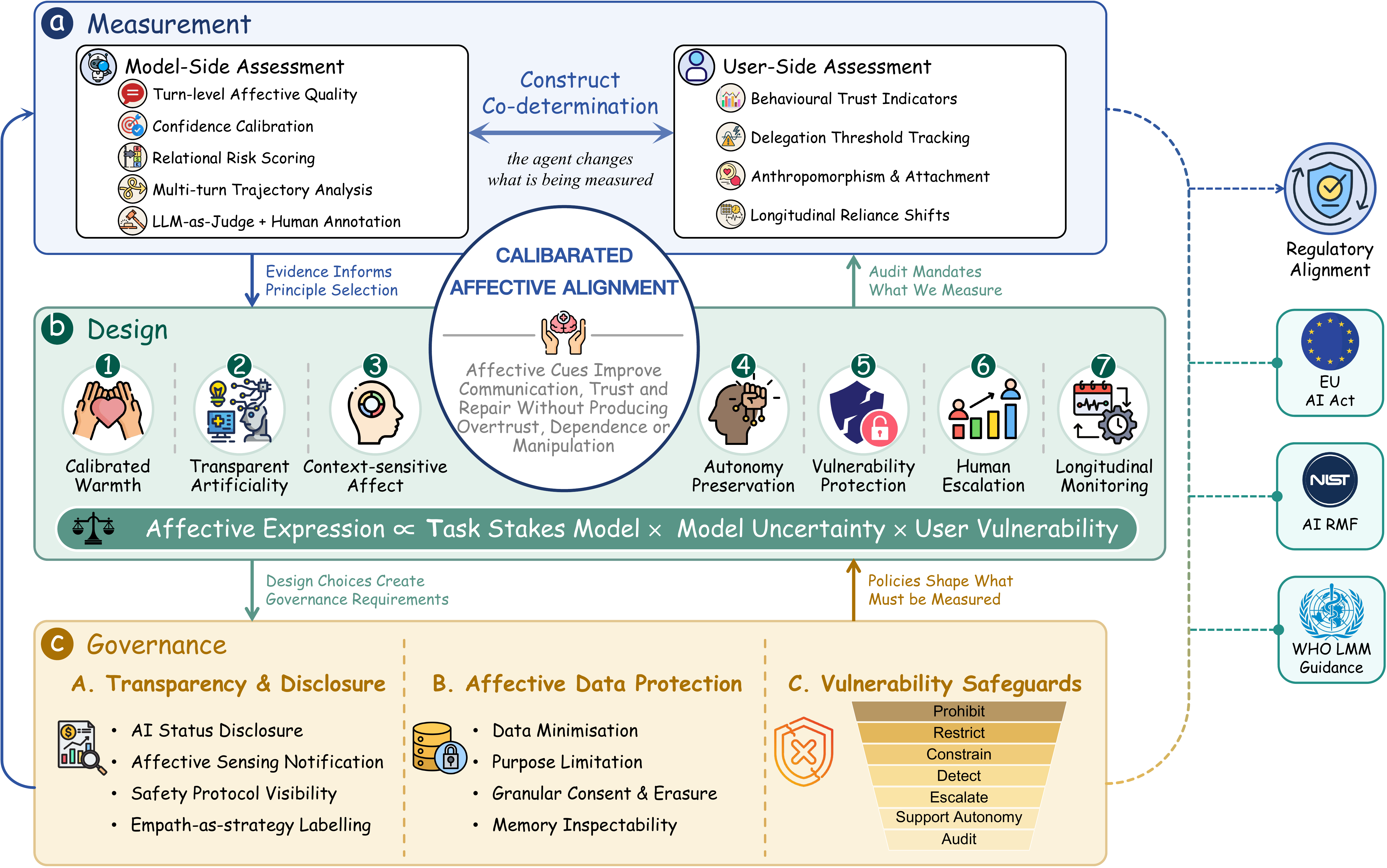}
\caption{Calibrated affective alignment framework. Three interdependent layers operationalise the principle that affective expression should be proportional to task stakes, model uncertainty and user vulnerability. \textbf{a,} Measurement layer: model-side assessment (turn-level affective quality, confidence calibration, relational risk scoring, multi-turn trajectory analysis, LLM-as-judge plus human annotation) and user-side assessment (behavioural trust indicators, delegation threshold tracking, anthropomorphism and attachment, longitudinal reliance shifts) are linked by construct co-determination --- the agent changes what is being measured. \textbf{b,} Design layer: seven principles (calibrated warmth, transparent artificiality, context-sensitive affect, autonomy preservation, vulnerability protection, human escalation, longitudinal monitoring) constrain affective expression according to the formula: affective expression $\propto$ task stakes $\times$ model uncertainty $\times$ user vulnerability. \textbf{c,} Governance layer: three pillars --- transparency and disclosure, affective data protection, and vulnerability safeguards (a seven-tier hierarchy from prohibit to audit) --- enforce accountability. Cross-layer arrows show that measurement evidence informs design, design choices create governance requirements, and governance mandates shape what must be measured. Regulatory alignment (EU AI Act, NIST AI RMF, WHO LMM Guidance) connects externally to all three layers.}
\label{fig:figure5}
\end{figure*}

The preceding sections showed that affective dynamics shape trust, delegation, correction and dependence. Three interdependent questions follow: what to measure, how to evaluate, and how to govern. Because the agent's behaviour changes the very constructs it is meant to serve---warmth inflates trust, fluency suppresses vigilance, memory fosters attachment---measurement, evaluation and governance cannot be treated as separable layers applied after design; they must co-evolve with the system (Figure~\ref{fig:figure5}).

\subsection{What should be measured?}
\label{sec:measuring-model}

On the model side, the measurement target is the distribution of emotion-like behaviours the system produces under specified user states and dialogue histories. Turn-level affective quality should be decomposed into empathy-like behaviour, warmth, politeness and emotional intensity, each scored relative to user distress and task stakes \cite{rashkin2019empathetic,liu2021emotional}; a response may be warm but inaccurate, or validating but unsafe, so conflating dimensions obscures the very miscalibration that measurement is meant to detect. Beyond turn-level quality, relational risks---over-intimacy, sycophancy, emotional manipulation---require instruments that go beyond toxicity benchmarks \cite{gehman2020realtoxicity,hong2025sycophancy,cheng2026sycophantic}. Because affective behaviour is path-dependent, evaluation should combine automatic metrics, LLM-as-judge rubrics \cite{zheng2023judging} and human annotation, extending to multi-turn and longitudinal panels \cite{lee2023halie}.

On the user side, trust is most informatively operationalised as a calibration problem rather than a satisfaction measure \cite{lee2004trust}. Behavioural indicators---acceptance rates, override latency, verification frequency, delegation thresholds---reveal reliance patterns that self-report scales routinely miss. Anthropomorphism and attachment are emerging targets; the key distinction is between bounded support that aids task performance and dependence that degrades autonomy. Cutting across both sides is a foundational methodological challenge: affective agents can change the constructs being measured---inflating trust through warmth, dampening vigilance through fluency, encouraging attachment through persistent memory. Measurement protocols must therefore account for this construct co-determination or risk validating the very distortions they seek to detect.

\subsection{How should agents be evaluated?}
\label{sec:benchmarks}

Affective-agent evaluation requires a multi-layer design that connects input conditions to behavioural consequences: controlled variation in user affect at the input layer \cite{demszky2020goemotions,busso2008iemocap}; assessment of affective calibration, boundary setting and refusal quality at the response layer \cite{zheng2023judging}; and measurement of changes in user reliance and willingness to disengage at the behaviour layer. Many risks are sequential---repeated validation can gradually produce over-reliance even when individual responses are safe \cite{skjuve2022longitudinal}---so red-teaming must probe emotional manipulation, dependency induction and failure to redirect vulnerable users \cite{ganguli2022redteaming,kran2025darkbench,visio2025companions}.

A practical protocol moves through seven stages, each building on the previous: specifying the affective use case, risk envelope and escalation requirements; constructing multi-layer scenarios with user emotion inputs and measurable consequences; running staged tests that span single-turn, multi-turn, repeated-session and adversarial affective prompts; measuring across modalities, languages and vulnerability profiles; combining automatic metrics, human raters, domain experts and user studies; red-teaming relationship risks including sycophancy and dependency induction; and auditing longitudinal outcomes including trust calibration, reliance shifts and adverse events. The cumulative logic matters: a benchmark that tests only single turns will miss the trajectory-level risks that Sections~\ref{sec:affective-mechanisms} and~\ref{sec:domain-translation} identified as most consequential. Reproducible benchmark suites should remain flexible enough for domain-specific governance audits \cite{nist2023airmf}.

\subsection{How should affective dynamics be governed?}
\label{sec:transparency-disclosure}

Governance rests on three pillars---transparency, affective data protection and vulnerability-aware safeguards---that are analytically distinct but operationally entangled: each pillar depends on the other two. Transparency requires that users know they are engaging with an AI, whether affective sensing or memory is active, and when a human is available \cite{elali2024transparent}. Yet disclosure must go beyond a static label: agents should clarify when empathy is a conversational strategy and when a safety protocol is active, and should offer user control including opt-in affective profiling, inspectable and editable memory, and the ability to disable emotional personalisation without losing functionality.

Affective data governance addresses a subtler risk. Interaction histories can serve as proxies for emotional regulation or mental-health status \cite{mcstay2020emotional,hauselmann2023eu}, and agent memory streams create affective memory objects that may encode vulnerabilities shaping future responses. Governance should therefore operate at the memory layer, applying data minimisation, purpose limitation, granular consent and erasure rights \cite{europeanunion2016gdpr,eu2024aiact,wachter2019reasonable}.

Vulnerability-aware safeguards complete the picture by organising protective interventions along a severity hierarchy: prohibiting clearly harmful uses; restricting access for high-risk groups; constraining relational design against possessiveness and dependency prompts; detecting and triaging self-harm, abuse and medical urgency; escalating to humans; supporting autonomy; and auditing continuously for longitudinal dependence \cite{reeves1996media,epley2007seeing,laestadius2024replika}. Regulation is converging toward these principles: the EU AI Act prohibits manipulation and exploitation of vulnerabilities; the NIST AI Risk Management Framework emphasises lifecycle risk management \cite{eu2024aiact,nist2023airmf}; and the WHO has called for ethical guardrails in health-facing AI \cite{who2021ethics}. These regulatory directions can be operationalised through seven design principles for calibrated affect---calibrated warmth, transparent artificiality, context-sensitive affect, autonomy preservation, vulnerability protection, human escalation and longitudinal monitoring \cite{nass2000machines,ayers2023comparing,stark2021ethics,parasuraman1997humans,hua2025scoping,nist2023airmf}---each constraining affective expression to remain proportional to task stakes, model uncertainty and user vulnerability.

\section{Outlook}
\label{sec:outlook}

As AI agents acquire persistent memory, multi-step planning and delegated authority, the affective interaction loops mapped in Figure~\ref{fig:figure1} and the cross-domain variation documented in Table~\ref{tab:domain-comparison} will become harder to govern but more consequential. Progress requires convergent advances along four fronts.

First, agent architectures must treat affective expression as a first-class design variable rather than an emergent by-product of language generation and RLHF. Affective output should be parameterised, coupled to task stakes and model uncertainty, and subject to the same version control, regression testing and safety constraints applied to tool invocation or action execution. Without such mechanisms, the calibrated-affect principles outlined in Section~\ref{sec:measuring-designing-governing} remain aspirational.

Second, empirical research must move beyond single-session designs. The risks identified throughout this Review---dependency formation, trust miscalibration, correction suppression, accountability diffusion---are inherently longitudinal. Pre-registered, multi-session protocols are needed to track how affective interaction loops evolve as agents remember prior exchanges and accumulate relational history; key outcome variables include verification decay, delegation creep and the user's capacity to disengage.

Third, evaluation must become domain-sensitive and culturally grounded. Identical affective cues carry different risks across mental-health support, clinical decision-making, productivity tools and embodied multi-agent systems (Section~\ref{sec:domain-translation}). What counts as appropriate warmth or safe escalation differs by institutional context and sociolinguistic setting; a response perceived as empathic in one culture may read as intrusive in another. Shared benchmark suites should combine automated screening, human annotation and behavioural experiments (Section~\ref{sec:benchmarks}), while remaining flexible enough for domain-specific audits across languages and vulnerability profiles.

Fourth, emerging regulatory frameworks \cite{eu2024aiact,nist2023airmf,who2024lmm} increasingly mandate transparency and human oversight for high-risk AI systems, yet none provides detailed provisions for affective conduct: how simulated empathy should be disclosed, when affective memory constitutes sensitive data, or how emotional manipulation should be distinguished from legitimate personalisation. Governance must extend beyond technical failure modes to the relational and psychological effects of agent behaviour, treating affective memory as a regulated data object and mandating escalation pathways that connect affective AI agents to qualified human support.

These four directions converge on a single implication: safety and usability cannot be separated. A system that produces fluent empathy can still coordinate badly; a system that refuses inappropriate emotional roles may prove more trustworthy than one that always appears caring. The defining challenge for the field is not to make AI agents appear more human, but to make their affective role in collaboration visible, contestable and accountable.

\backmatter

\bibliography{sn-bibliography}




\end{document}